\documentclass[aps,prl,reprint,superscriptaddress,showpacs,floatfix]{revtex4-2}

\usepackage{amsmath}
\usepackage{amssymb}
\usepackage{graphics}
\usepackage{graphicx}
\usepackage{epstopdf}
\usepackage{dcolumn} 
\usepackage{bm}
\usepackage{color}
\usepackage{xcolor}
\usepackage[colorlinks=true, linkcolor=blue, citecolor=blue, urlcolor=blue]{hyperref}

\newcommand{\BDelta}{\boldsymbol{\Delta}}

\newcommand{\Bkappa}{\boldsymbol{\kappa}}
\newcommand{\BLambda}{\boldsymbol{\Lambda}}

\begin{document}

\title{Geometric phase-space nonseparability triggers giant optical shifts}

\author{Kaiqi Zhu}
\affiliation{School of Physical Science and Technology, Soochow University, Suzhou 215006, China}
\affiliation{Suzhou Key Laboratory of Intelligent Photoelectric Perception, Soochow University, Suzhou 215006, China}

\author{Yonglei Liu}
\affiliation{School of Physical Science and Technology, Soochow University, Suzhou 215006, China}
\affiliation{Suzhou Key Laboratory of Intelligent Photoelectric Perception, Soochow University, Suzhou 215006, China}

\author{Yao Zhao}
\affiliation{School of Physical Science and Technology, Soochow University, Suzhou 215006, China}
\affiliation{Suzhou Key Laboratory of Intelligent Photoelectric Perception, Soochow University, Suzhou 215006, China}

\author{Zhongyi Hu}
\affiliation{School of Physical Science and Technology, Soochow University, Suzhou 215006, China}
\affiliation{Suzhou Key Laboratory of Intelligent Photoelectric Perception, Soochow University, Suzhou 215006, China}

\author{Jiahui Shen}
\affiliation{School of Physical Science and Technology, Soochow University, Suzhou 215006, China}
\affiliation{Suzhou Key Laboratory of Intelligent Photoelectric Perception, Soochow University, Suzhou 215006, China}

\author{Yimeng Zhu}
\affiliation{School of Physical Science and Technology, Soochow University, Suzhou 215006, China}
\affiliation{Suzhou Key Laboratory of Intelligent Photoelectric Perception, Soochow University, Suzhou 215006, China}

\author{Lin Liu}
\affiliation{School of Physical Science and Technology, Soochow University, Suzhou 215006, China}
\affiliation{Suzhou Key Laboratory of Intelligent Photoelectric Perception, Soochow University, Suzhou 215006, China}

\author{Yangjian Cai}
\thanks{yangjiancai@sdnu.edu.cn}
\affiliation{Shandong Provincial Key Laboratory of Light Field Manipulation Physics and Applications \& School of Physics and Optoelectronics, Shandong Normal University, Jinan 250358, China}

\author{Fei Wang}
\thanks{fwang@suda.edu.cn}
\affiliation{School of Physical Science and Technology, Soochow University, Suzhou 215006, China}
\affiliation{Suzhou Key Laboratory of Intelligent Photoelectric Perception, Soochow University, Suzhou 215006, China}

\author{Sergey A.~Ponomarenko}
\thanks{serpo@dal.ca}
\affiliation{Department of Electrical and Computer Engineering, Dalhousie University, Halifax, Nova Scotia, B3J 2X4, Canada}
\affiliation{Department of Physics and Atmospheric Science, Dalhousie University, Halifax, Nova Scotia, B3H 4R2, Canada}

\author{Yahong Chen}
\thanks{yahongchen@suda.edu.cn}
\affiliation{School of Physical Science and Technology, Soochow University, Suzhou 215006, China}
\affiliation{Suzhou Key Laboratory of Intelligent Photoelectric Perception, Soochow University, Suzhou 215006, China}

\begin{abstract}
Nonseparability among multiple degrees of freedom has enabled fundamental advances in structured light and related applications. Here we unveil a previously overlooked form of nonseparability in phase space, which we term geometric phase-space nonseparability. The latter arises solely from the wavefront curvature of a conventional wave packet, such as a fundamental Gaussian beam. This phase-space structure manifests as a position-dependent transverse-momentum distribution across the beam profile leading to the giant spatial and angular beam shifts upon reflection at a planar interface that we  predict analytically and observe experimentally. Remarkably, the curvature-induced phase-space correlation remains robust against spatial-coherence degradation, allowing the giant shifts to persist even in the nearly incoherent regime. Our results establish wavefront curvature as a general mechanism for engineering beam shifts across optical, acoustic, and matter-wave systems.
\end{abstract}

\maketitle

\textit{Introduction.---}Optical fields possess multiple degrees of freedom that can become intrinsically coupled \cite{Shen22a}. The resulting nonseparability, often referred to as classical entanglement of light \cite{Forbes19a,Paneru20,Qin26}, underlies a broad class of structured optical fields \cite{Forbes21a,He22}, including spatiotemporal wave packets \cite{Yessenov22,Zhan24,Hebri26} and vector beams \cite{Forbesvector}. Beyond these familiar forms of nonseparability, the phase-space representation of light \cite{Alonso11} reveals intrinsic correlations between real-space position and transverse momentum of photons composing optical wave packets. Recent studies have shown that both fully coherent \cite{Chowdhury13,Prabhakar15} and partially coherent  \cite{Ponomarenko21b,Jiang25a,PSA26} wave packets endowed with orbital angular momentum exhibit phase-space nonseparability. By tailoring their orbital-angular-momentum-related phase-space structure, giant and controllable beam shifts can be generated at planar interfaces \cite{Chen25, Bashiri26}.

\begin{figure}[h!]
\centering
\includegraphics[width=\linewidth]{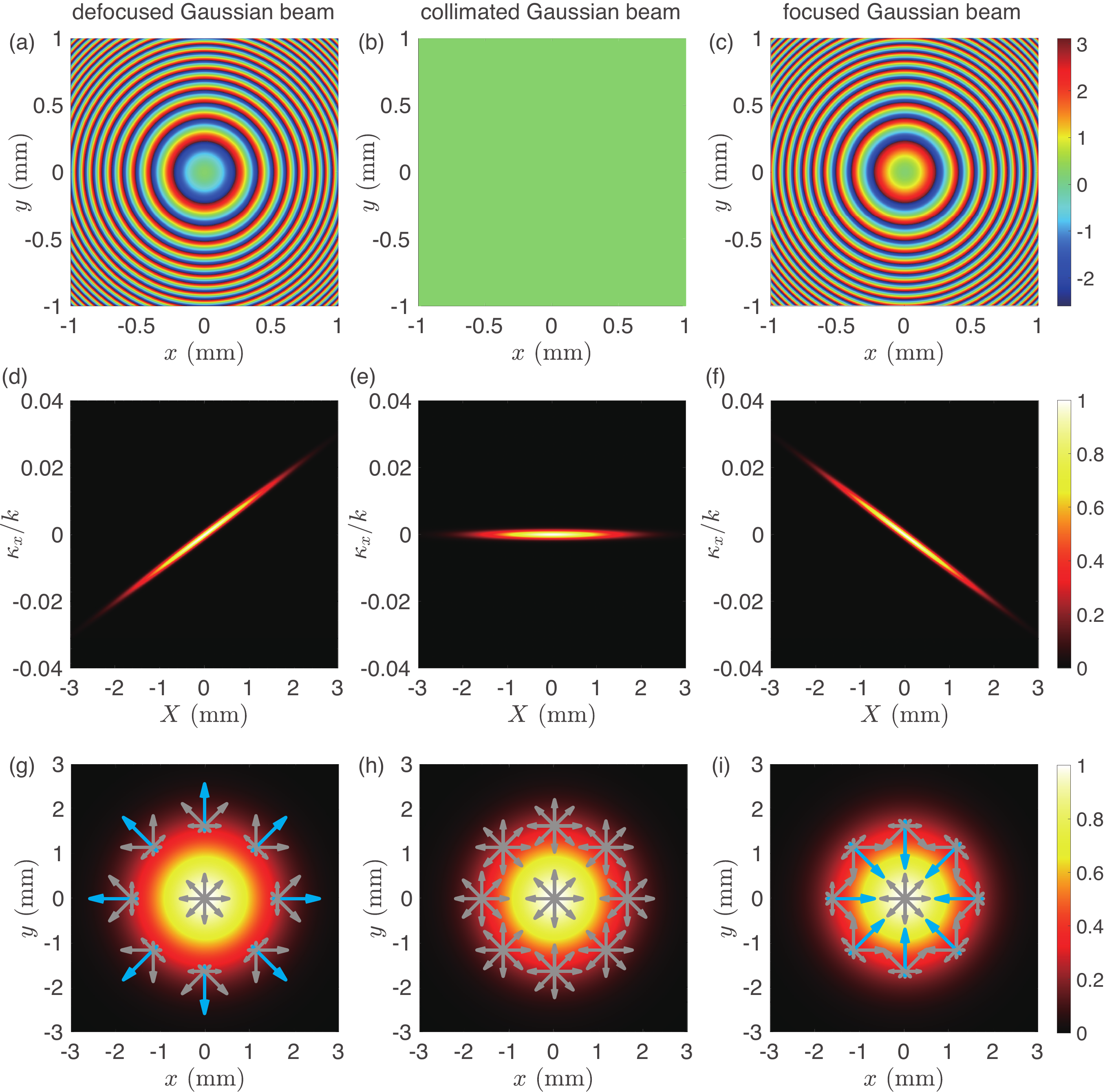}
\caption{Geometric phase-space nonseparability of Gaussian Schell-model beams with curved wavefronts. (a)--(c) Phase distributions of defocused ($f=-100~\mathrm{mm}$), collimated ($f \to \infty$), and focused ($f=100~\mathrm{mm}$) Gaussian beams. (d)--(f) Corresponding projections of the Wigner distribution functions onto $(X,\kappa_x)$ plane for $\sigma_I = 1~\mathrm{mm}$ and $\sigma_c = 0.1~\mathrm{mm}$. (g)--(i) Spatial distributions of transverse wave vectors across the Gaussian beam profile, with the local mean transverse wave vectors highlighted by blue arrows. The spatially inhomogeneous transverse momentum distribution reveals the underlying phase-space nonseparability of the beam.}
\label{fig1}
\end{figure}

Here, we identify a fundamentally different route to phase-space nonseparability that requires no orbital angular momentum. Even a fundamental Gaussian beam with a curved wavefront possesses a nonseparable phase-space geometry, manifested as a position-dependent transverse-momentum distribution across the beam profile (cf.~Fig.~\ref{fig1}). We demonstrate theoretically and experimentally that this geometric phase-space nonseparability triggers giant in-plane Goos--H\"anchen (GH) and out-of-plane Imbert--Fedorov (IF) shifts upon reflection at a simple air--glass interface. Remarkably, the wavefront-curvature-induced phase-space correlation remains robust against spatial-coherence degradation, allowing the giant shifts to persist even for nearly incoherent wave packets.

Wavefront curvature arises ubiquitously during propagation, diffraction, focusing, and defocusing, yet it has not previously been recognized as a direct mechanism for generating giant beam shifts upon reflection. Existing approaches typically rely on resonant or structured interfaces \cite{Schreier98,Soboleva12,Wan20}, ultrahigh-order modes \cite{Wu19,Dai20}, or beams carrying orbital angular momentum \cite{Chen25, Bashiri26, Bliokh09,Merano10}. In contrast, our approach relies on the wavefront curvature of a fundamental Gaussian beam and is applicable to unstructured dielectric interfaces. Therefore, we establish wavefront curvature as a simple and ubiquitous tool to control position--momentum correlations and beam shifts across a wide spectrum of physical wave systems.

\textit{Geometric phase-space nonseparability.---}To account for the effect of optical coherence in a general framework, we consider a generic partially coherent Gaussian Schell-model beam \cite{Mandelbook} endowed with a curved wavefront, imparted by a thin lens with focal length $f$. We can express the corresponding Wigner distribution function~\cite{BastiaansRev} as (see Sec.~S1 of \cite{SM} for the derivation)
\begin{align}
\label{WDFGSM}
\!\!\! \mathcal{W}(\mathbf{R},\Bkappa)
 \propto 
\exp \left( - \frac{\mathbf{R}^2}{2 \sigma_I^2} \right)
\exp \left[ -\frac{\sigma_\mathrm{eff}^2}{2}  
\left( \Bkappa+\frac{k}{f} \mathbf{R}\right)^2 \right],
\end{align}
where $\mathbf{R} = (X,Y)$ and $\Bkappa = (\kappa_x,\kappa_y)$  represent the position and momentum vectors in phase space, with $\Bkappa$ corresponding to the transverse wave vector; $k=\omega/c$ is the wave number, and $\sigma_\mathrm{eff}^{-2} = \sigma_c^{-2} + (2 \sigma_I)^{-2}$. Here, $\sigma_I$ and $\sigma_c$ denote the beam width and transverse spatial coherence width, respectively. In the limit $\sigma_c \to \infty$, the beam reduces to a fully coherent Gaussian beam.

It follows from Eq.~(\ref{WDFGSM}) that the curved wavefront intrinsically induces a non-separable phase-space structure: The Wigner distribution function does not factorize into functions of $\mathbf{R}$ and $\Bkappa$. This phase-space geometry is governed by the wavefront curvature, characterized by $f$.

Figure~\ref{fig1} visualizes this geometric phase-space nonseparability. It manifests as a tilt of the Wigner distribution, whose orientation is determined by the sign of the wavefront curvature. The Wigner distributions of focused and defocused Gaussian beams tilt in opposite directions, whereas that of a collimated Gaussian beam with a planar wavefront is untilted and separable in phase space. This tilted phase-space geometry gives rise to a spatially inhomogeneous transverse-momentum distribution, as illustrated in Figs.~\ref{fig1}(g) and \ref{fig1}(i). Specifically, the local mean transverse wave vectors, indicated by the blue arrows, vary radially across the beam profile. For $f<0$ ($f>0$), they point outward (inward), corresponding to a defocused (focused) beam. By contrast, for the beam with a planar wavefront, the transverse wave vectors remain uniformly distributed, as shown in Fig.~\ref{fig1}(h).

To quantify the degree of phase-space nonseparability, we perform a Schmidt decomposition of the Wigner distribution function and evaluate the Schmidt rank (see Sec.~S2 of \cite{SM})
\begin{equation}\label{K}
   K=\frac{z_R^2/f^2+1/\xi_c^2+1/4}{1/\xi_c^2+1/4}\geq 1,
\end{equation}
where $z_R = k\sigma_I^2$ is the Rayleigh range associated with the beam width $\sigma_I$, and $\xi_c=\sigma_c/\sigma_I$ is a dimensionless coherence parameter. Nonseparability implies $K>1$. The analysis of Eq.~(\ref{K}) reveals that if the wavefront is sufficiently curved ($f \ll z_R$), then $K\gg1$, indicating strong nonseparability for coherent $\xi_c\gg 1$, partially coherent $\xi_c\sim 1$, and nearly incoherent $1\ll 1/\xi_c\ll z_R/f$ source wave packets. In all of these cases, the spatial coherence width $\sigma_c$ affects only the width of the momentum distribution, leaving the underlying phase-space correlation unchanged. This description breaks down only for extremely incoherent source beams (Secs. S2 and S3 of \cite{SM}).

\begin{figure}[b!]
\centering
\includegraphics[width=0.9\linewidth]{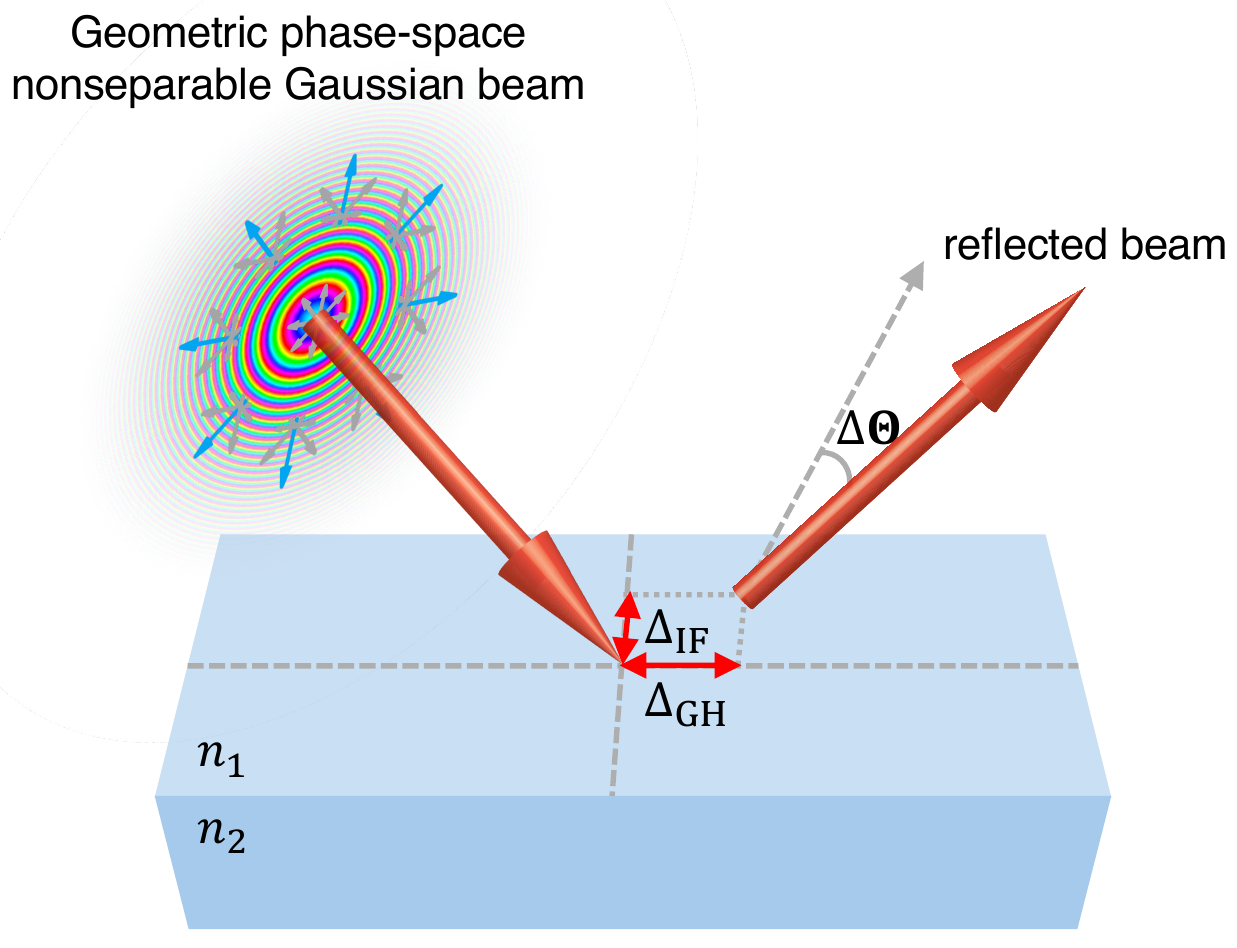}
\caption{Reflection of a Gaussian beam with a curved wavefront at an interface between two dielectric media with refractive indices $n_1$ and $n_2$. The arrows across the incident beam profile illustrate the spatially inhomogeneous transverse momentum distribution generated by the curved wavefront, with the mean transverse wave vectors highlighted in blue. Upon partial reflection, the reflected beam exhibits giant spatial GH and IF shifts, denoted by $\Delta_\mathrm{GH}$ and $\Delta_\mathrm{IF}$, together with an angular deviation $\Delta\boldsymbol{\Theta}$ arising from the geometric phase-space structure of the incident beam.}
\label{fig2}
\end{figure}

\textit{Giant beam shifts.---}The beam-reflection geometry is shown in Fig.~\ref{fig2}. A uniformly polarized Gaussian Schell-model beam with a curved wavefront is incident at an angle $\theta$ on a planar interface separating two transparent, nonmagnetic media with refractive indices $n_1$ and $n_2$. Using a general phase-space formalism \cite{Chen25}, we obtain the centroid of the reflected beam at a distance $z$ from the interface as (Sec.~S4 of \cite{SM})
\begin{equation}
\label{centroid}
\mathbf{R}_c(z) = \BDelta + \frac{2 z_R}{f} \BLambda - \frac{2 z}{z_R} \left(\frac{1}{4} + \frac{1}{\xi_c^2} + \frac{z_R^2}{f^2} \right) \BLambda, 
\end{equation}
where $\BDelta$ and $\BLambda$ are two-component vectors describing the in-plane and out-of-plane contributions to the beam shift. Their explicit expressions in terms of the angle-dependent complex Fresnel reflection coefficients for $p$ and $s$ polarizations and the polarization components of the incident beam are given in Sec.~S4 of \cite{SM}.

The first term in Eq.~(\ref{centroid}) represents the intrinsic spatial GH and IF shifts of a fundamental Gaussian beam with a planar wavefront and is therefore independent of the phase-space geometry. The second term arises from the geometric phase-space nonseparability induced by wavefront curvature and depends explicitly on the focal length $f$. It vanishes as $f \to \infty$, when the wavefront becomes planar. The third term describes the propagation-dependent angular shifts, which increase linearly with $z$. The three factors $1/4$, $1/\xi_c^2$, and $z_R^2/f^2$ represent, respectively, the intrinsic angular shift of a Gaussian beam, the contribution of finite spatial coherence, and the contribution of geometric wavefront curvature.

Several conclusions follow directly from Eq.~(\ref{centroid}). First, when the wavefront is sufficiently curved, namely for small $|f|$, the curvature-induced spatial and angular shifts can dominate over the intrinsic beam shifts, thereby producing giant beam shifts. Second, the sign of the curvature-induced spatial shift is set by the sign of $f$, so focused and defocused beams exhibit opposite shifts. By contrast, wavefront curvature changes only the magnitude of the angular shift, not its sign.

Next, spatial coherence affects only the angular shifts and leaves the spatial shifts unchanged, allowing giant spatial shifts to persist even for nearly incoherent light. Moreover, when the wavefront curvature is sufficiently large, the geometric term proportional to $z_R^2/f^2$ dominates both the intrinsic and coherence-dependent angular contributions. The angular shifts then become effectively insensitive to spatial coherence. Finally, for a lossless interface, the vector $\BLambda$ is nonzero only under partial reflection. Therefore, the curvature-induced giant beam shifts reported here occur only under partial reflection.

We can explain the physical origin of the wavefront-curvature-triggered beam shifts as follows. The wavefront curvature of the incident beam tilts the Wigner distribution, establishing a transverse position--momentum correlation that increases with the curvature (reduced $|f|$). The momentum-dependent Fresnel response acts as a filter upon wave packet reflection. The position--momentum correlation converts this momentum selectivity into an asymmetric spatial weighting across the reflected beam, thereby producing a large centroid displacement. Concurrently, the curvature-induced broadening of the transverse-momentum distribution allows this momentum selectivity to amplify imbalances among angular components, enhancing the angular shift. For a planar wavefront, these curvature-induced effects vanish because the Wigner distribution is separable in position and momentum.

\textit{Experimental results.---}
We test these predictions by considering Gaussian beam reflection from an air--glass prism interface, with the wavefront curvature controlled using convex and concave lenses of different focal lengths. Near the Brewster angle, the reflected $p$-polarized beam can be strongly reshaped and displaced, whereas the $s$-polarized beam remains nearly Gaussian with negligible centroid displacement. Hence, we determine the beam shifts from the difference between the reflected-beam centroids for $p$- and $s$-polarized incidence. This differential scheme avoids detector recalibration as the prism is rotated and suppresses systematic errors. Further experimental details are provided in Sec.~S5 of \cite{SM}.

\begin{figure}[b!]
\centering
\includegraphics[width=0.85\linewidth]{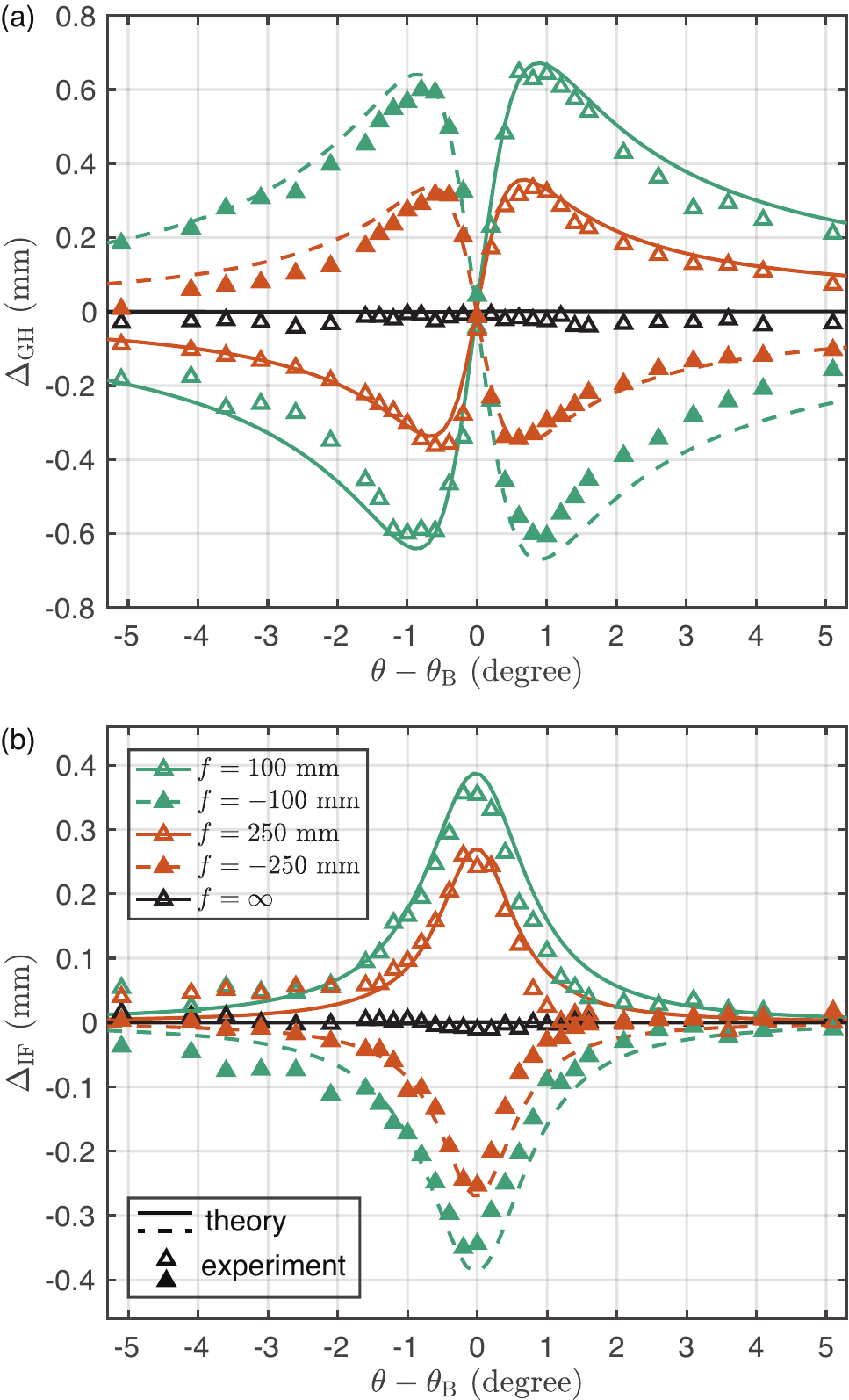}
\caption{Spatial beam shifts induced by geometric phase-space nonseparability under partial reflection. (a) Spatial GH shift $\Delta_\mathrm{GH}$ and (b) IF shift $\Delta_\mathrm{IF}$ for fully coherent Gaussian beams carrying curved wavefronts generated by focusing ($f>0$) and defocusing ($f<0$) lenses, shown as functions of $\theta-\theta_\mathrm{B}$. Here $\theta_\mathrm{B}=\arctan(n_2/n_1)=56.7456^\circ$ denotes the Brewster angle and $\theta$ is the angle of incidence. The markers represent the experimental measurements, while the solid and dashed curves show the theoretical predictions obtained from Eq.~(\ref{centroid}). The case $f \to \infty$ corresponds to a collimated Gaussian beam with a planar wavefront.}
\label{fig3}
\end{figure}

Figure~\ref{fig3} shows the measured spatial GH and IF shifts for fully coherent Gaussian beams, together with the predictions of Eq.~(\ref{centroid}). Both shifts increase rapidly as $|f|$ decreases and reverse sign when the wavefront curvature is reversed. For $f=\pm100~\mathrm{mm}$, the IF shift reaches approximately $0.4~\mathrm{mm}$ ($750$ wavelengths) at the Brewster angle, while the GH shift reaches approximately $0.65~\mathrm{mm}$ ($1220$ wavelengths) near $\theta-\theta_{\mathrm B}\simeq\pm0.8^\circ$. By contrast, for a planar wavefront, both shifts vanish under partial reflection and remain on the order of a few wavelengths under total internal reflection \cite{Merano09, Bliokh13a}. The measurements agree well with the analytical predictions, demonstrating giant and controllable beam shifts at a simple planar interface.

\begin{figure}[h!]
\centering
\includegraphics[width=0.85\linewidth]{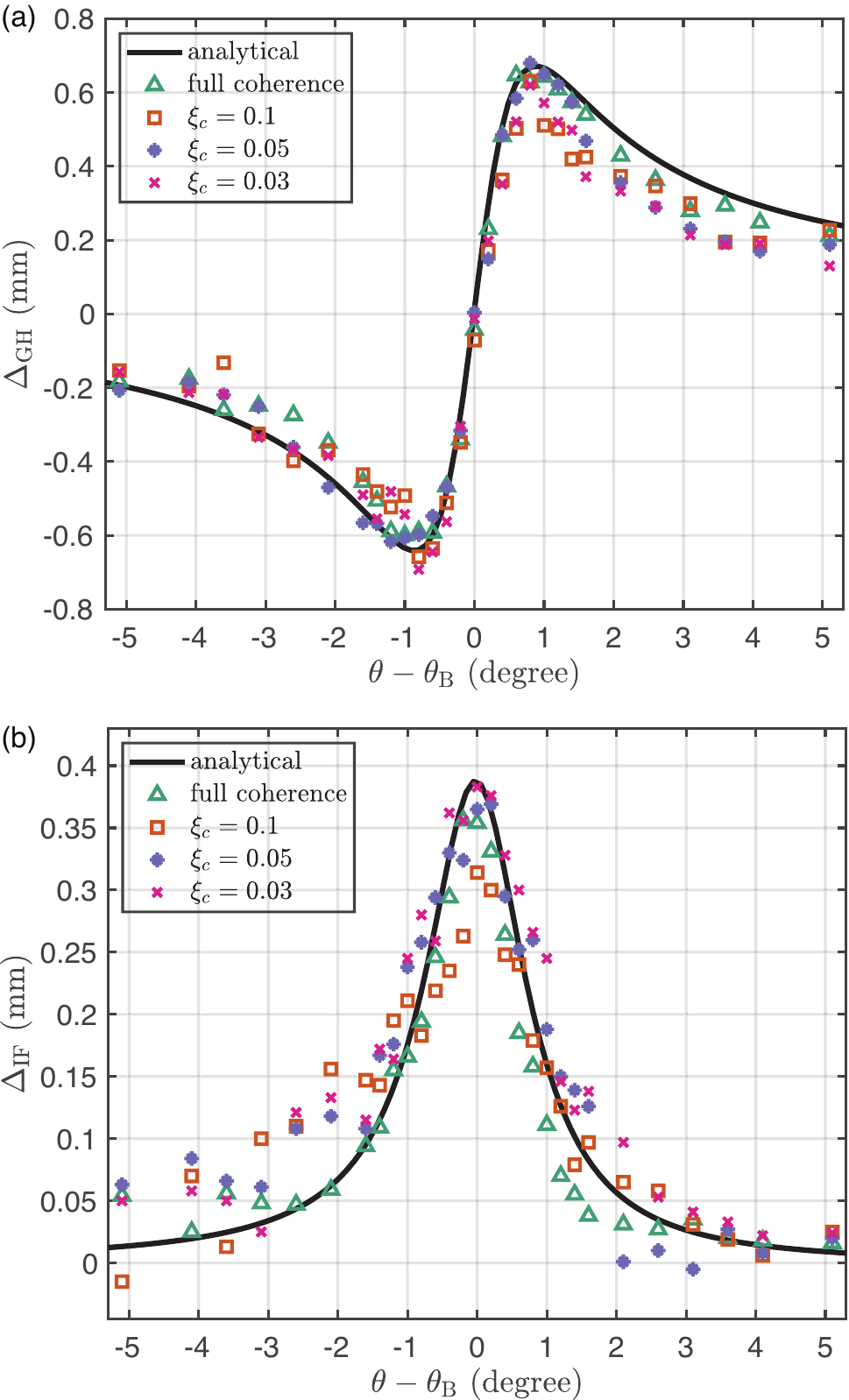}
\caption{Effect of spatial coherence on the spatial beam shifts. (a) Spatial GH shift $\Delta_\mathrm{GH}$ and (b) IF shift $\Delta_\mathrm{IF}$ for partially coherent Gaussian Schell-model beams of different state of spatial coherence characterized by the coherence parameter $\xi_c$. All incident beams have the same wavefront curvature corresponding to $f = 100~\mathrm{mm}$. The markers represent the experimental measurements, while the solid curves show the theoretical predictions obtained from Eq.~(\ref{centroid}).}
\label{fig4}
\end{figure}

We next examine the role of spatial coherence using Gaussian Schell-model beams. Figure~\ref{fig4} compares the measured and predicted shifts for $\xi_c=0.1$, $0.05$, and $0.03$, with the beam width fixed at $\sigma_I=1~\mathrm{mm}$ and the focal length at $f=100~\mathrm{mm}$. Both the GH and IF shifts remain nearly unchanged as the spatial coherence decreases, and giant shifts are still directly observable at $\xi_c=0.03$. Reduced coherence broadens the Wigner distribution transverse to its curvature-induced tilt but leaves the underlying position--momentum correlation largely intact (see Fig.~S2 of \cite{SM}), explaining the observed robustness.

\begin{figure}[t!]
\centering
\includegraphics[width=\linewidth]{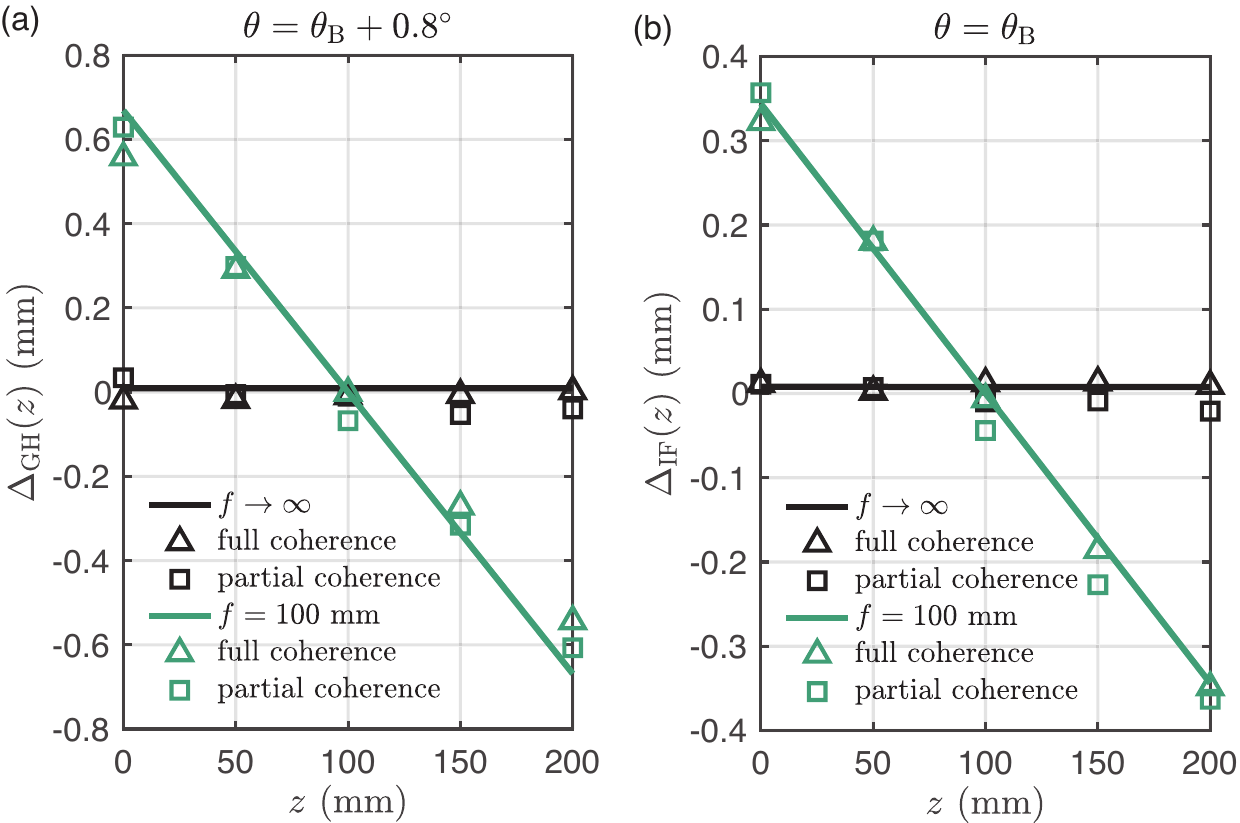}
\caption{Giant angular beam shifts induced by geometric wavefront curvature. (a) GH shift and (b) IF shift as functions of the propagation distance $z$ for a fully coherent Gaussian beam and a Gaussian Schell-model beam with $\xi_c = 0.03$. The markers represent the experimental measurements, while the solid curves show the theoretical predictions obtained from Eq.~(\ref{centroid}). Black markers and curves correspond to beams with planar wavefronts, whereas green markers and curves correspond to beams with curved wavefronts corresponding to $f = 100~\mathrm{mm}$.}
\label{fig5}
\end{figure}

Finally, we measure the curvature-enhanced angular beam shifts. Figure~\ref{fig5} shows the GH and IF shifts as functions of the propagation distance $z$, together with the analytical predictions of Eq.~(\ref{centroid}). Their linear dependence on $z$ directly reveals the angular deflection of the reflected beam. For $f=100~\mathrm{mm}$ and $\sigma_I=1~\mathrm{mm}$ (see the green curves and markers in Fig.~\ref{fig5}), the wavefront curvature of the beam enhances the angular shifts by a factor of approximately $5.58\times10^4$ relative to those of a flat-wavefront Gaussian beam, whose angular deflection is consequently negligible (see the black curves and markers in Fig.~\ref{fig5}). The measurements also show that the angular shifts remain nearly unchanged as the spatial coherence is reduced to $\xi_c=0.03$. This robustness follows from Eq.~(\ref{centroid}): under the present conditions, the curvature contribution $z_R^2/f^2$ substantially exceeds the coherence-dependent contribution $1/\xi_c^2$, so the angular shifts are dominated by wavefront curvature.

\textit{Conclusions.---}We have revealed geometric phase-space nonseparability as a previously overlooked property of wave packets with curved wavefronts. This nonseparable structure manifests as a position-dependent transverse-momentum distribution across the beam profile. We have predicted theoretically and observed experimentally  giant spatial and angular beam shifts in reflection triggered by geometric phase-space nonseparability. We have shown that wavefront curvature controls both the magnitude and direction of the spatial shifts and can enhance the angular shifts by several orders of magnitude. Remarkably, the curvature-induced phase-space correlation remains robust against spatial-coherence degradation, allowing the giant shifts to persist even for nearly incoherent wave packets. 

Our results establish wavefront curvature as a simple and powerful mechanism for giant beam shift engineering through geometric phase-space structure manipulation. Compared with previous approaches involving interface structure design or endowing incident beams with orbital angular momentum, our method is experimentally much simpler and, most importantly, naturally applicable to low-coherence optical wave packets, including sunlight, X-rays, and LEDs. Moreover, geometric phase-space nonseparability can be imposed on acoustic wave packets with acoustic lenses~\cite{al1,al2,al3} and matter wave packets, including cold atoms and electrons, with the aid of, for instance, magnetic~\cite{ml} or laser (dipole)~\cite{JOSAB91} lenses.~More broadly, our findings reveal that ordinary wavefront curvature can encode a hidden geometric phase-space structure in otherwise conventional Gaussian wave packets, opening new avenues for precision metrology, beam engineering, and wave-based~sensing~technologies.

\textit{Acknowledgments.---}This work was supported by the National Natural Science Foundation of China (12274310, 12404348, 12347114, 12274311, 12192254, 92250304, W2441005, 12534014), the National Key Research and Development Project of China (2022YFA1404800), the Natural Science Foundation of Shandong Province (ZR2025ZD21), and the Natural Sciences and Engineering Research Council of Canada (RGPIN-2025-04064).


\begin{thebibliography}{99}

\bibitem{Shen22a}
Y. Shen and C. Rosales-Guzm\'an,
Nonseparable states of light: From quantum to classical,
Laser Photonics Rev. \textbf{16}, 2100533 (2022).

\bibitem{Forbes19a}
A. Forbes, A. Aiello, and B. Ndagano,
Classically entangled light,
Prog. Opt. \textbf{64}, 99 (2019).

\bibitem{Paneru20}
D. Paneru, E. Cohen, R. Fickler, R. W. Boyd, and E. Karimi,
Entanglement: Quantum or classical?,
Rep. Prog. Phys. \textbf{83}, 064001 (2020).

\bibitem{Qin26}
Y. Qin, Z. Wang, Q. Liu, and X. Fu,
Quantum--classical analogy in structured light,
Appl. Phys. Rev. \textbf{13}, 021326 (2026).

\bibitem{Forbes21a}
A. Forbes, M. de Oliveira, and M. R. Dennis,
Structured light,
Nat. Photonics \textbf{15}, 253 (2021).

\bibitem{He22}
C. He, Y. Shen, and A. Forbes,
Towards higher-dimensional structured light,
Light Sci. Appl. \textbf{11}, 205 (2022).

\bibitem{Yessenov22}
M. Yessenov, L. A. Hall, K. L. Schepler, and A. F. Abouraddy,
Space-time wave packets,
Adv. Opt. Photon. \textbf{14}, 455 (2022).

\bibitem{Zhan24}
Q. Zhan,
Spatiotemporal sculpturing of light: A tutorial,
Adv. Opt. Photon. \textbf{16}, 163 (2024).

\bibitem{Hebri26}
D. Hebri and S. A. Ponomarenko,
Space-time optics,
Prog. Opt. \textbf{71}, 37 (2026).

\bibitem{Forbesvector}
C. Rosales-Guzm\'an, B. Ndagano, and A. Forbes,
A review of complex vector light fields and their applications,
J. Opt. \textbf{20}, 123001 (2018).

\bibitem{Alonso11}
M. A. Alonso,
Wigner functions in optics: Describing beams as ray bundles and pulses as particle ensembles,
Adv. Opt. Photon. \textbf{3}, 272 (2011).

\bibitem{Chowdhury13}
P. Chowdhury, A. S. Majumdar, and G. S. Agarwal,
Nonlocal continuous-variable correlations and violation of Bell's inequality for light beams with topological singularities,
Phys. Rev. A \textbf{88}, 013830 (2013).

\bibitem{Prabhakar15}
S. Prabhakar, S. G. Reddy, A. Aadhi, C. Perumangatt, G. K. Samanta, and R. P. Singh,
Violation of Bell's inequality for phase-singular beams,
Phys. Rev. A \textbf{92}, 023822 (2015).

\bibitem{Ponomarenko21b}
S. A. Ponomarenko,
Twist phase and classical entanglement of partially coherent light,
Opt. Lett. \textbf{46}, 5958 (2021).

\bibitem{Jiang25a}
H. Jiang, P. Peng, H. Wang, X. Peng, Y. Chen, F. Wang, Y. Cai, and L. Liu,
Observation of the classical entanglement of twisted partially coherent light in phase space,
Opt. Lett. \textbf{50}, 7071 (2025).

\bibitem{PSA26}
S. A. Ponomarenko and M. Hajati,
Wigner picture of partially coherent accelerating beams,
Opt. Lett. \textbf{51}, 2144 (2026).

\bibitem{Chen25}
Y. Chen and S. A. Ponomarenko,
Phase-space nonseparability, partial coherence, and optical beam shifts,
Phys. Rev. Lett. \textbf{135}, 193801 (2025).

\bibitem{Bashiri26}
S. Bashiri, Y. Chen, and S. A. Ponomarenko,
Engineering the geometric optics limit of wave-packet reflection from a planar interface,
Phys. Rev. A \textbf{114}, L011501 (2026).

\bibitem{Schreier98}
F. Schreier, M. Schmitz, and O. Bryngdahl,
Beam displacement at diffractive structures under resonance conditions,
Opt. Lett. \textbf{23}, 576 (1998).

\bibitem{Soboleva12}
I. V. Soboleva, V. V. Moskalenko, and A. A. Fedyanin,
Giant Goos--H\"anchen effect and Fano resonance at photonic crystal surfaces,
Phys. Rev. Lett. \textbf{108}, 123901 (2012).

\bibitem{Wan20}
R. Wan and M. S. Zubairy,
Tunable and enhanced Goos--H\"anchen shift via surface plasmon resonance assisted by a coherent medium,
Opt. Express \textbf{28}, 6036 (2020).

\bibitem{Wu19}
F. Wu, J. Wu, Z. Guo, H. Jiang, Y. Sun, Y. Li, J. Ren, and H. Chen,
Giant enhancement of the Goos--H\"anchen shift assisted by quasibound states in the continuum,
Phys. Rev. Applied \textbf{12}, 014028 (2019).

\bibitem{Dai20}
H. Dai, L. Yuan, C. Yin, Z. Cao, and X. Chen,
Directly visualizing the spin Hall effect of light via ultrahigh-order modes,
Phys. Rev. Lett. \textbf{124}, 053902 (2020).

\bibitem{Bliokh09}
K. Y. Bliokh, I. V. Shadrivov, and Y. S. Kivshar,
Goos--H\"anchen and Imbert--Fedorov shifts of polarized vortex beams,
Opt. Lett. \textbf{34}, 389 (2009).

\bibitem{Merano10}
M. Merano, N. Hermosa, J. P. Woerdman, and A. Aiello,
How orbital angular momentum affects beam shifts in optical reflection,
Phys. Rev. A \textbf{82}, 023817 (2010).

\bibitem{Mandelbook}
L. Mandel and E. Wolf,
\textit{Optical Coherence and Quantum Optics}
(Cambridge University Press, Cambridge, England, 1995).

\bibitem{BastiaansRev}
M. J. Bastiaans,
Application of the Wigner distribution function to partially coherent light,
J. Opt. Soc. Am. A \textbf{3}, 1227 (1986).

\bibitem{SM}
See Supplemental Material for details on the derivations of Eqs.~(\ref{WDFGSM}), (\ref{K}), and (\ref{centroid}), the effect of optical coherence on the geometric phase-space nonseparability, and the experimental details. The Supplemental Material includes Refs.~\cite{Forbes19a, Ponomarenko21b, Chen25, Meh, Morse, Chan02}.

\bibitem{Meh}
E. Merzbacher,
\textit{Quantum Mechanics}
(Wiley \& Sons, 1998).

\bibitem{Morse}
P. M. Morse and H. Feshbach,
\textit{Methods of Theoretical Physics}
(Technology Press, 1946).

\bibitem{Chan02}
K. W. Chan, C. K. Law, and J. H. Eberly,
Localized single-photon wave functions in free space,
Phys. Rev. Lett. \textbf{88}, 100402 (2002).

\bibitem{Merano09}
M. Merano, A. Aiello, M. P. van Exter, and J. P. Woerdman,
Observing angular deviations in the specular reflection of a light beam,
Nat. Photonics \textbf{3}, 337 (2009).

\bibitem{Bliokh13a}
K. Y. Bliokh and A. Aiello,
Goos--H\"anchen and Imbert--Fedorov beam shifts: An overview,
J. Opt. \textbf{15}, 014001 (2013).

\bibitem{al1}
J. T. Welter, S. Sathish, D. E. Christensen, P. G. Brodrick, J. D. Heebl, and M. R. Cherry,
Focusing of longitudinal ultrasonic waves in air with an aperiodic flat lens,
J. Acoust. Soc. Am. \textbf{130}, 2789 (2011).

\bibitem{al2}
G. Y. Song, B. Huang, H. Y. Dong, Q. Cheng, and T. J. Cui,
Broadband focusing acoustic lens based on fractal metamaterials,
Sci. Rep. \textbf{6}, 35929 (2016).

\bibitem{al3}
Y. Tang and E. S. Kim,
Ring-focusing Fresnel acoustic lens for long depth-of-focus focused ultrasound with multiple trapping zones,
J. Microelectromech. Syst. \textbf{29}, 692 (2020).

\bibitem{ml}
A. El-Kareh,
\textit{Electron Beams, Lenses, and Optics}
(Elsevier, Amsterdam, 2012).

\bibitem{JOSAB91}
G. M. Gallatin and P. L. Gould,
Laser focusing of atomic beams,
J. Opt. Soc. Am. B \textbf{8}, 502 (1991).

\end{thebibliography}
\end{document}